\documentclass[conference]{IEEEtran}
\usepackage{cite}
\usepackage{amsmath,amssymb,amsfonts}
\usepackage{algorithmic}
\usepackage[ruled,vlined, linesnumbered]{algorithm2e}
\usepackage{graphicx}
\usepackage{textcomp}
\usepackage{xcolor}
\usepackage{soul}
\usepackage{braket}
\def\BibTeX{{\rm B\kern-.05em{\sc i\kern-.025em b}\kern-.08em
    T\kern-.1667em\lower.7ex\hbox{E}\kern-.125emX}}
\begin{document}

\title{A Shuttle-Efficient Qubit Mapper for Trapped-Ion Quantum Computers
\vspace{-4mm}
{\footnotesize \textsuperscript{}}
\thanks{}
}

\newtheorem{example}{Example}[section]
\makeatletter 
\newcount\SOUL@minus
\makeatother  
\author{\IEEEauthorblockN{\textsuperscript{} }
\IEEEauthorblockA{\textit{} \\
\textit{}\\
\\
}
\and
\IEEEauthorblockN{Suryansh Upadhyay}
\IEEEauthorblockA{\textit{School of EECS} \\
\textit{Pennsylvania State University}\\
University Park, PA \\
sju5079@psu.edu}
\and
\IEEEauthorblockN{Abdullah Ash Saki}
\IEEEauthorblockA{\textit{Zapata Computing} \\
MA, USA \\
axs1251@psu.edu}
\and
\IEEEauthorblockN{Rasit Onur Topaloglu}
\IEEEauthorblockA{\textit{IBM Corporation} \\
Hopewell Junction, NY \\
rasit@us.ibm.com}
\and
\IEEEauthorblockN{Swaroop Ghosh}
\IEEEauthorblockA{\textit{School of EECS} \\
\textit{Pennsylvania State University}\\
University Park, PA \\
szg212@psu.edu}
\and
}

\maketitle
\thispagestyle{plain}
\pagestyle{plain}

\begin{abstract}
Trapped-ion (TI) quantum computer is one of the forerunner quantum technologies. However, TI systems can have a limited number of qubits in a single trap. Execution of meaningful quantum algorithms requires a multiple trap system. In such systems, the computation may frequently involve ions from two different traps for which the qubits must be co-located in the same trap, hence one of the ions needs to be shuttled (moved) between traps, increasing the vibrational energy, degrading fidelity, and increasing the program execution time. The choice of initial mapping influences the number of shuttles. The existing Greedy policy counts the number of gates occurring between each pair of qubits and assigns edge weight. The qubits with high edge weights are placed close to each other. However, it neglects the stage of the program at which the gate is occurring. Intuitively, the contribution of the late-occurring gates to the initial mapping reduces since the ions might have already shuttled to a different trap to satisfy other gate operations. In this paper, we target this gap and propose a new policy especially for programs with considerable depth and high number of qubits (valid for practical-scale quantum programs). Our policy is program adaptive and prioritizes the gates re-occurring at the initial stages of the program over late occurring gates. Our technique achieves an average reduction of 9\% shuttles/program (with 21.3\% at best) for 120 random circuits and enhances the program fidelity up to 3.3X (1.41X on average).

\end{abstract}


\section{Introduction}
Quantum Computing (QC) exploits phenomena such as superposition, entanglement, and interference to efficiently explore exponentially large state spaces and compute solutions for certain classically intractable problems. Application domains of QC include machine learning~\cite{b1}, security~\cite{b3}, drug discovery~\cite{b4}, computational quantum chemistry~\cite{b5} and optimization~\cite{b6}. 
On one hand, researchers are proposing new quantum algorithms to speed up computation, while on the other hand, various technologies like superconducting, trapped-ion (TI), and photonics are also being studied to design efficient quantum bits or qubits.
Two of the forerunner qubit technologies are superconducting qubits~\cite{b7} and trapped-ion (TI) qubits~\cite{b9, b10, b11}.
\vspace{3pt}


TI qubit offers several advantages such as perfectly identical qubits 
, long coherence times, and all-to-all connectivity among qubits~\cite{b12}, making them one of the most promising technology candidates for building NISQ devices.
Several companies such as IonQ and Honeywell are pursuing this technology. Recently, Honeywell reported a trapped-ion system with a high quantum volume of 1024\cite{b14}. Quantum volume is a metric introduced by IBM to measure the overall capability and performance of a quantum computing system regardless of technology\cite{b15}. Various TI hardware systems~\cite{b16, b19} are also commercially available through quantum cloud services such as Honeywell, AWS Braket, and Microsoft Azure. 
\vspace{3pt}

A major hurdle in realizing large TI systems is confining many ions in a single trap as it decreases the spacing between ions, making it challenging to pulse a qubit using laser controllers selectively. Moreover, the gate time becomes slow, which results in longer program execution time. Therefore, the pathway to scalability in TI systems involves multiple interconnected traps.
However, in a multi-trap system, computation is sometimes required on data from ions situated in different traps. For such cases, one ion needs to be shuttled (moved) from one trap to another so that the ions are co-located and the gate operation can be performed. A compiler adds shuttle operations to a quantum program to satisfy the inter-trap communication, and the shuttle operation increases program execution time and degrades quantum gate fidelity. Therefore, it is essential to minimize the number of shuttle operations.
Fig. \ref{2}a shows a simplified 2-trap system interconnected by a shuttle path. 
Here, we assume that each trap can accommodate a maximum of 4 ions, i.e., trap capacity equals 4 per trap. 
\vspace{3pt}

To scale up TI technology for near-term applications, a modular and scalable architecture called the Quantum Charge Coupled Device (QCCD) is proposed~\cite{b22}. 
An extensive architectural study for multi-trap trapped ion systems has been reported in~\cite{b23} where a compiler and a simulator has been developed for such systems with experimentally calibrated values. 
The initial mapping entails the assignment of logical qubits from the program to the traps and the relative position of qubits inside a trap. It is essential to have an optimum initial mapping as it influences the total number of shuttles during the program execution. For example, if we randomly map 2 logical qubits from the program with the highest number of gates between them in different traps, there will be an increase of shuttles between the two traps compared to the case when they were mapped together in the same trap. A good initial mapping policy thus tries to place ions with frequent gates together so that communication can be minimized. 
\vspace{3pt}

A Greedy mapping policy ~\cite{b23} has been proposed for the initial mapping \emph(elaborated with an example in Section III) where a graph is created for the program with qubits as nodes and gates between qubits as edges. An edge weight is assigned to a qubit pair depending on the frequency of occurrence for the gate between them. It maps the edges in the descending order of the weight to place ions with frequent gates together. However, it does not consider the gate position in the program and program-specific parameters like the number of qubits, gates, and depth. For instance, considering a program with 1000 2-qubit gates, a gate occurring at the 900\textsuperscript{th} instance intuitively, won’t contribute much to the initial mapping (ions might have already shuttled to a different trap to satisfy other gate operations) as the one occurring at the 10\textsuperscript{th} instance. 
\vspace{3pt}

Various initial qubit mapping heuristics have been extensively reported for superconducting qubits ~\cite{b27} ~\cite{b28}, however literature lacks work on TI qubit mapping policy which is fundamentally different; this is a technical gap we target in this paper. Superconducting qubits tend to interact only with their nearest neighbors, whereas trapped ions can easily interact with several other ions but only if they are within the same trap, and through costly shuttles if they are in a different trap. For superconducting qubit systems, qubit-mapping algorithms explicitly attempt to minimize the amount of swap gates employed, also considering other parameters such as two-qubit gate error rates, single-qubit error rates, and execution time. On the other thand for TI qubits, as their operations are  less prone to errors ~\cite{b29} and qubits have better connectivity within the trap, the mapping heuristics aim to minimize the number of shuttles between the traps.

\vspace{3pt}
In this paper, we present an efficient initial qubit mapping heuristic to target the inefficiency in the Greedy mapping policy. Our basic idea is to attenuate the weight of late occurring gates in the program. We explore three different edge weight allotment heuristics, i.e., step, linear, and decay. For example, the linear edge weight policy reduces the weight of the gate linearly with increasing number of stages. Based on the intuition developed from the above explorations, we propose an optimized initial qubit mapper based on a penalizing approach which assigns negative weight to the late occurring gates. The proposed policy considers program-specific parameters such as, number of qubits, number of gates, and depth, making it more custom for the program.
\vspace{3pt}

The rest of the paper is organized as follows: Section II describes the background for quantum computing and TI systems. Section III presents the proposed initial mapper heuristics. Section IV compares the Greedy mapper with the proposed heuristics. Finally, we conclude in Section V. 

\section{Background}

In this section, we discuss the basics of a quantum computer, TI systems, and the terminologies used in this paper.

\subsection{Qubits}

Qubits are the building block of a quantum computer that store data (i.e., $\ket{0}$ and $\ket{1}$) as various internal states. Due to quantum superposition property, a qubit can be in both $\ket{0}$ and $\ket{1}$ simultaneously unlike a classical bit which can either be 0 or 1. Hence, an n-qubit system can represent all $2^n$ basis states simultaneously rather than a classical n-bit register in exactly one of the $2^n$ states. A qubit state is represented as a $\psi = a \ket{0} + b \ket{1}$ where $a$ and $b$ are complex probability amplitudes of states $\ket{0}$ and $\ket{1}$, respectively. Measurement collapses the qubit to one state and returns classical bits 0 or 1 with probabilities $|a|^2$ and $|b|^2$, respectively.

\subsection{Quantum Gates}
\begin{figure}
    \centering
    \includegraphics[width= 3.5in]{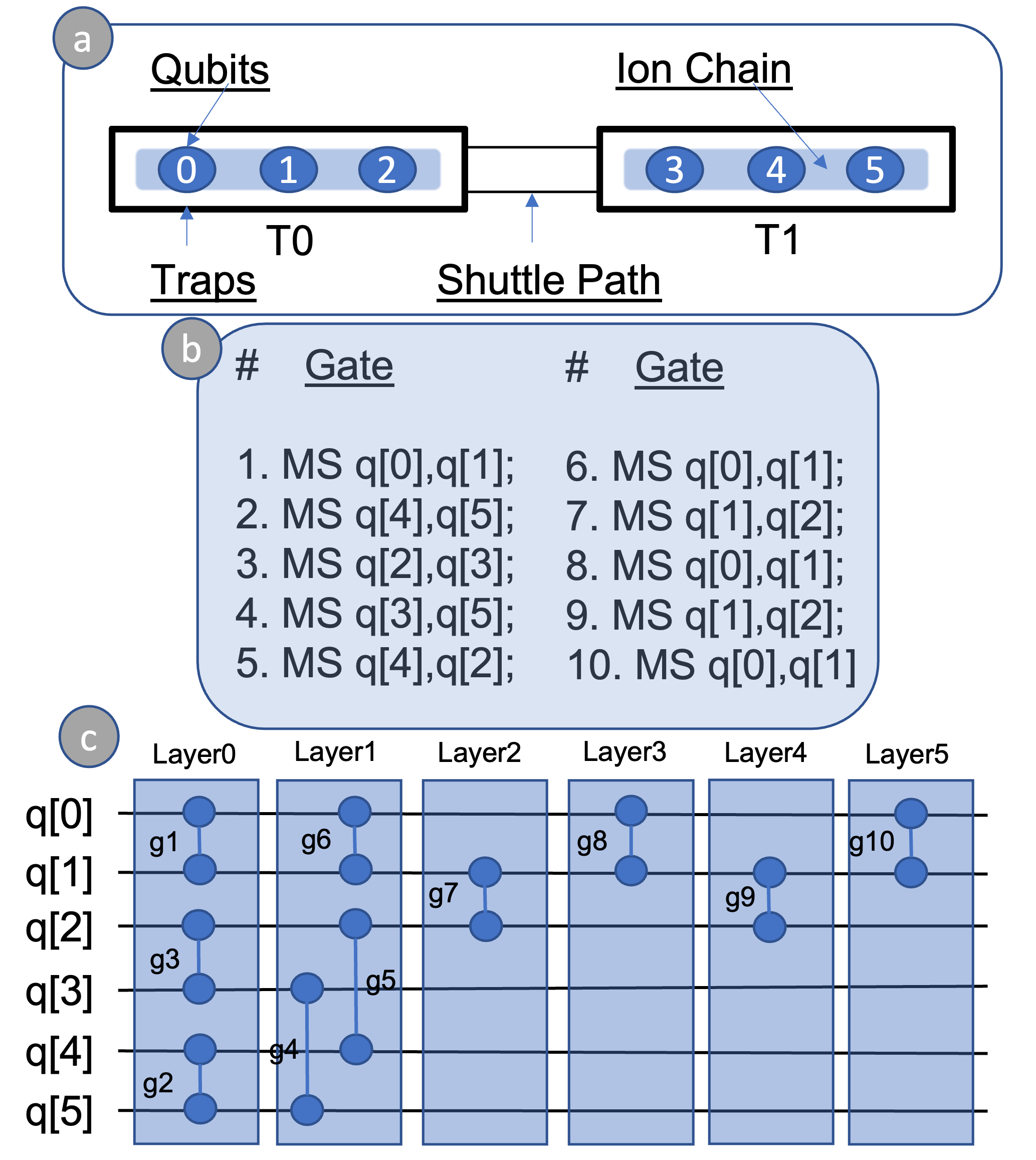}
    \caption{(a) A two trap TI system with three qubits each. (b)  A sample quantum program consisting of ten 2-qubit Mølmer–Sørensen (MS)~\cite{b24, b26} gates. (c) Gate dependency graph of the sample program. 
    }
    \label{2}
\vspace{-4mm}
\end{figure}

Computation in quantum systems is performed by manipulating the information stored in the qubits. 
The gates transform the qubit amplitudes to get the desired output. Gates are realized using pulses such as radio frequency (RF) and laser pulses. The gates are reversible and are mathematically represented by unitary matrices. In a QC program, a sequence of gates is executed on a set of appropriately initialized qubits. At present, physically realized gates in TI systems are 1-qubit and 2-qubit. For example, the IonQ TI system offers 1-qubit GPI, GPI2, and GZ gates and 2-qubit Mølmer–Sørensen (MS)~\cite{b24, b26} gate. Fig. \ref{2}(b) illustrates a program with 2-qubit MS gates.

\subsection{Gate Fidelity}
Quantum gates in existing quantum computers are erroneous due to imperfect qubit control, errors in pulse implementation, and external interference. Therefore, the gates have finite and non-negligible error rates $(\epsilon)$ during execution. The gate fidelity ($F$) is usually defined as the complement of the error rate. A lower gate fidelity will introduce more errors in the output and can completely decimate the result. 


\subsection{Gate Dependency Graph and Depth of a Program}
A quantum program can be converted to a gate dependency graph, i.e., a directed acyclic graph (DAG). The dependency graph consists of layers. Gates in a layer are independent but depend on one or more gates from the previous layers. The program's depth is obtained from the number of layers of the gate dependency graph. For example, Fig. \ref{2}c shows the dependency graph for the sample program in Fig. \ref{2}b. Gates $g5$ and $g6$ are independent of each other in Layer 1 as they operate on different sets of qubits. However, $g5$ and $g6$ depend on $g1$ and $g3$, respectively, which means the former gates cannot be executed before $g1$ and $g3$. The depth for the sample program is six, which is calculated from the number of layers.

\subsection{Trapped Ion QC Systems}
\subsubsection{Traps and Ion Chain}
Trapped ion QC system are implemented by trapping ionized atoms like $Yb$ or $Ca$ between electrodes using electromagnetic field~\cite{b19}. 
Data $\ket{0}$ and $\ket{1}$ are encoded as internal states such as hyper-fine or Zeeman states of the ions. Using Fig. \ref{2}a, we illustrate various components of a 2-trap TI system. The ions are organized in form of an ion chain inside a trap. \emph{Trap capacity} is the maximum number of ions that a trap can accommodate. The traps are connected by a shuttle path which allows movement (shuttle) of an ion from one trap to another if needed. During the initial allocation of ions, a part of the total trap capacity is loaded with ions and the remaining capacity (termed as \emph{communication capacity}) is kept unoccupied to allow for shuttled ions from other traps.
\vspace{3pt}


\subsubsection{Shuttle Operation}
\begin{figure}
    \centering
    \includegraphics[width=3.2in]{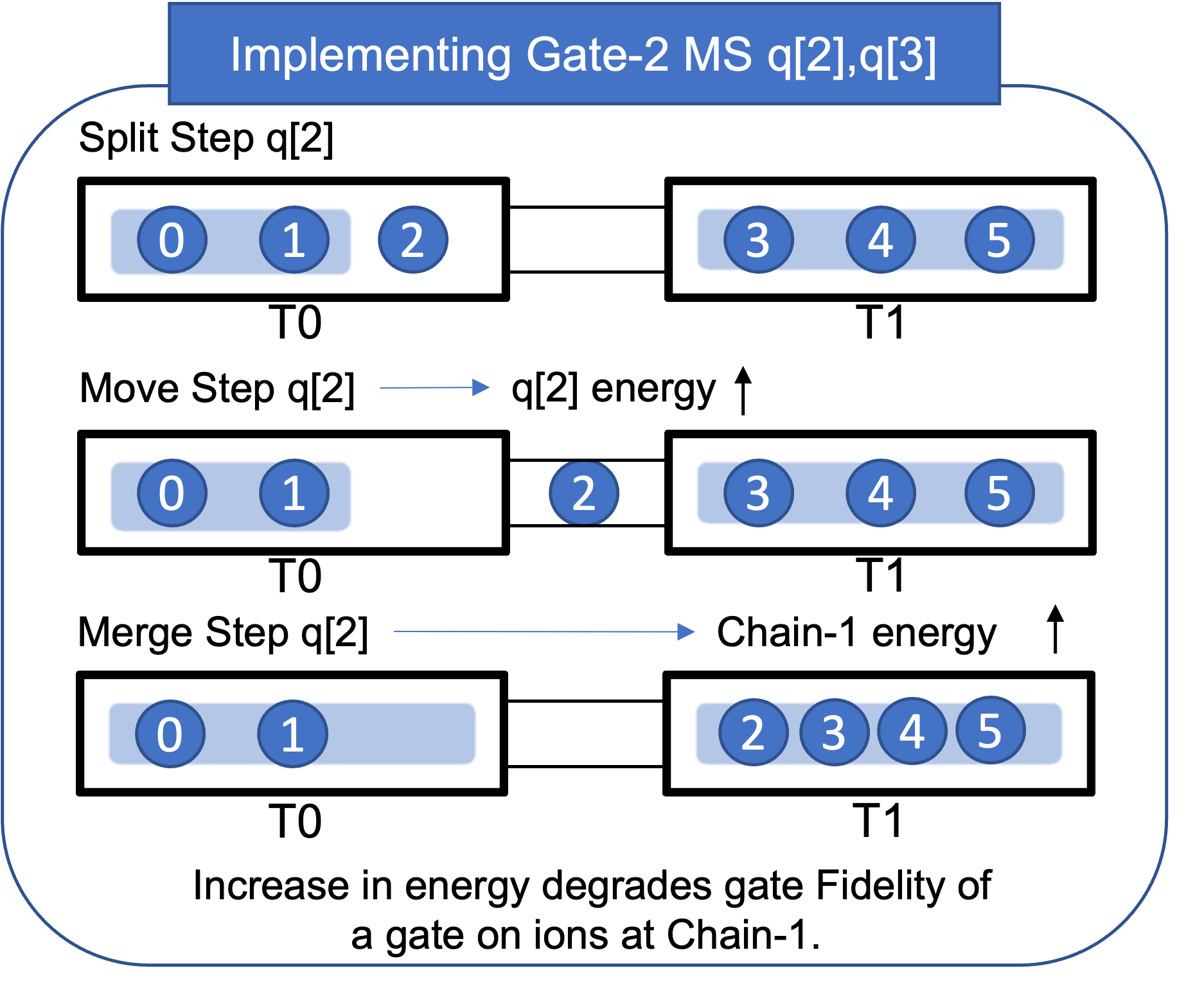}
    \caption{Shuttle steps to move ion-2 from trap T0 to trap T1.}
    \label{3}
\vspace{-4mm}
\end{figure}

A 2-qubit gate between ions from different traps requires a shuttle. For example, in Fig. \ref{2}b, the $1^{st}$ gate in the program MS q[0], q[1] involves ions from the same trap (T0) and can be executed directly. However, the $3^{rd}$ gate MS q[2], q[3] involves ions from different traps.
Therefore, a shuttle operation is needed to bring both ions into the same trap. For the shuttle operation (Fig. \ref{3}), first, ion-2 is split from Chain–0 and shuttled from T0 to T1, adding energy to the ion. Then, ion-2 is merged to the Chain-1. Finally, gate MS q[2],q[3] can be executed as the ions are in the same trap (T1). The merge operation increases the vibrational energy $(\bar{n})$ of chain-1 affecting the gate fidelity as follows:
\begin{equation}\label{eq:fidelity}
    F = 1 - \Gamma \tau - A (2 \bar{n} + 1) 
\end{equation}
where, ($\tau$) is gate time, ($\Gamma$) is trap heating rate  and ($\bar{n}$) is the vibrational energy. 
With increased vibrational energy $\bar{n}$, the subsequent gate operations on chain-1 will experience lower gate fidelities  per Eq.~\ref{eq:fidelity}.


\subsection{Initial Qubit Mapping}
The initial mapping refers to mapping of logical qubits to physical traps and their relative positions inside a trap for the first time. It is essential to have an optimal mapping logic as initial mapping influences the total number of shuttles. For example, the program qubits (0 to 5) from the sample program in Fig. \ref{2}b are initially mapped as $T0$: [0, 1, 2], $T1$: [3, 4, 5] considering a Greedy allocation \cite{b23, b24}
(detailed below). The program execution will start with this allocation and the mapping will be updated based on the number of shuttles.
\emph{In this work, we list the limitation of the Greedy mapper and propose a heuristic to reduce the net number of shuttles for various benchmarks compared to the Greedy mapper.}
\vspace{3pt}


\subsubsection{Greedy Mapping Policy}\label{AA}

In the Greedy policy, the qubits are mapped to the physical traps by considering the number of gates between the qubit pair. A quantum program can be modeled as a graph where each node represents a qubit, and an edge between two qubits represents a 2-qubit gate. Thus, the edge weight represents the number of 2-qubit gates between a pair of qubits. The Greedy policy maps the edges in the descending order of the weight, placing edges with high weight first, allowing qubits with a high number of gates close together, i.e., in the same trap. 
 The basic weight assignment algorithm for Greedy policy is illustrated in \emph{Algorithm 1}.It starts by placing the highest weighted edge in one of the traps. Next, for each edge with one mapped and one unmapped endpoint, the algorithm maps the unmapped qubit to the adjacent position, minimizing the total distance between the qubit and its neighbors.
The process is repeated for each unmapped edge in the descending order of the edge-weight. The Greedy mapping policy can be explained using the sample program in Fig. \ref{2}b. The edge weights of the program are, wt(0,1) = 4 (as the MS q[0],q[1] gate appears 4 times throughout the program), wt(1,2) = 2, and wt(4,5) = wt(2,3) = wt(3,5) = wt(4,2) = 1. Therefore, ions 0 and 1 are allocated first followed by ion 2, and finally, ions 3, 4, and 5.


\begin{algorithm}
\SetAlgoLined
\KwIn{gate map}
\KwOut{edge weights}

\For{gate $\in$ gate map}{
    \uIf{qubit\_edge\_weight $\in$ edge register}{
        qubit\_edge\_weight += 1\;
    }
    \Else{
        qubit\_edge\_weight = 1\;
    }
    update $edge~register$ with qubit\_edge\_register\;
}
\caption{Allocation of edge weight using the greedy policy}
\label{4}
\end{algorithm}

\subsubsection{Limitations of Greedy mapper}
The Greedy policy assigns a constant edge weight for every occurrence of a gate between two qubits regardless of the position of the gate in the program. A gate appearing at the beginning of the program is assigned the same edge weight as the one appearing at the end. In programs with considerable depth and many qubits, the initial mapping using this policy is not optimum. Hence, leaves the scope of further improvement in the number of shuttles.
Intuitively, the gates appearing toward the end of a deep program should not determine the initial mapping since the qubits might already have shuttled to a different trap.

\section{Optimized Initial mapper Heuristics}

\subsection{Basic Idea}
We propose a mapping heuristic where edge weights are assigned based on the number of gates between qubits and considering the stage of the program where the gate appears. We also consider program-specific parameters such as, the number of qubits, depth of the program \emph{(derived from the gate dependency graph)}, and the total number of gates while assigning edge weights. Priority is given to the gates re-occurring at the earlier stages of the program.


\subsection{Edge Weight Function}

We propose to improve the logic by assigning edge weights for any re-occurrence of a gate using a function rather than a constant value. However, the first occurrence of a gate in the program will be assigned the same constant value similar to the Greedy mapper policy. This process can be illustrated using the \emph{Algorithm 2}. Various types of decaying functions explored in this paper are explained in the following section. Here \emph{cnt} is just a variable used to implement a counter loop; it will take values from 0 to \emph{(number of gates in the program)} - 1. For the first occurrence of a gate, edge weight will be assigned a constant value\emph{(equal to the total number of gates in the program)}, followed by an increment in edge weight per the decaying function being used for any re-occurrence of the gate later in the program. We explain this algorithm using a sample program and a sample function in Example \ref{example1}. 
\vspace{3pt}

Our initial compilations are carried out using three different functions. By studying the effect of these functions on the number of shuttles, we arrive at our optimized initial mapping heuristic discussed later in this section.
\begin{algorithm}
\SetAlgoLined
\KwIn{gate map}
\KwOut{edge weights}

\For{gate $\in$ gate map}{
    \If{cnt $\leq$ \# of gates}{
        \uIf{qubit\_edge\_weight $\in$ edge register}{
            qubit\_edge\_weight += decaying function\{f(cnt)\}\;
            cnt += 1\;
        }
        \Else{
            qubit\_edge\_weight = total \# of gates\;
            cnt += 1\;
        }
    }
    update $edge~register$ with qubit\_edge\_register\;
}
\caption{Optimized edge weight heuristic}
\label{5}
\end{algorithm}
\subsubsection{The Decaying Step Function}
\begin{figure}[htbp]
\centerline{\includegraphics[width= 3.5in]{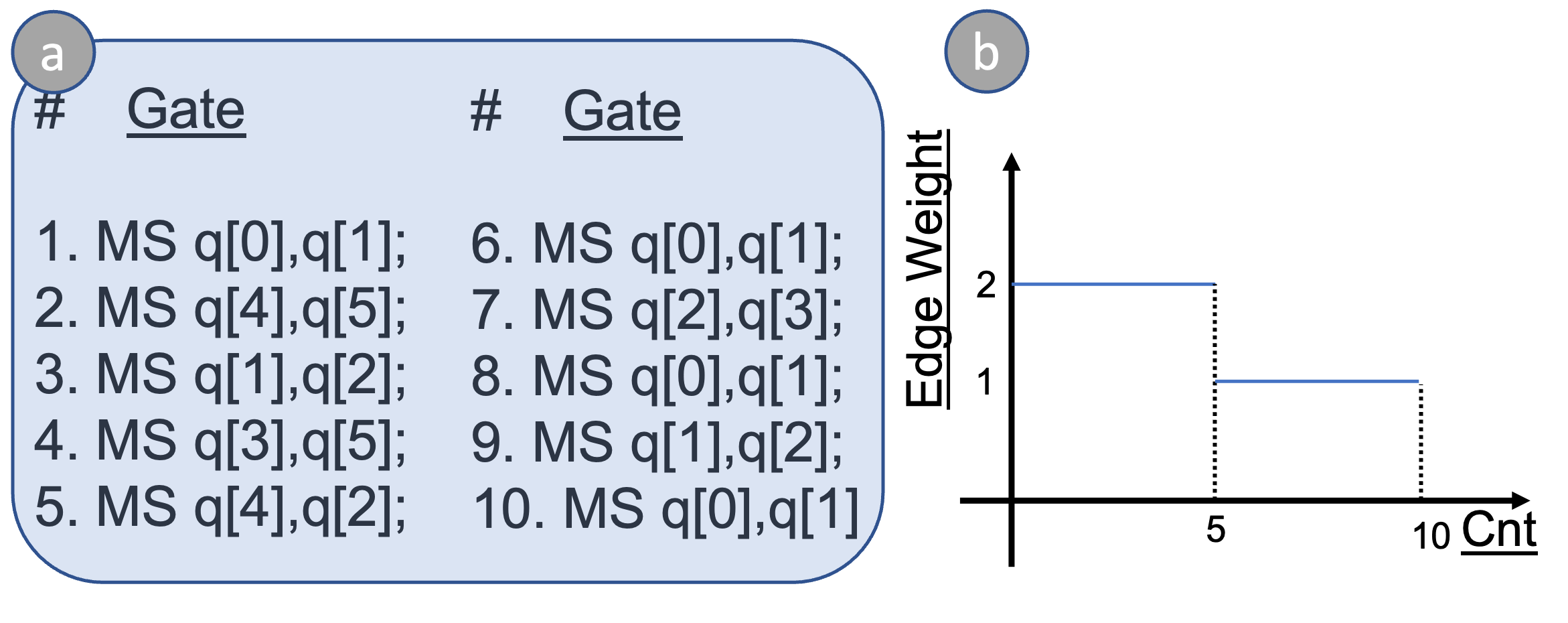}}
\caption{(a) A sample quantum program consisting of ten 2-qubit gates. (b)
A decaying step function for the sample program.}
\label{6}
\end{figure}

We divide our program into $n$ ($n$ being an empirical parameter) equal blocks for simplicity. For each gate in a block, we assign a constant edge weight. The edge weight is assigned to the block in the form of a decaying staircase function, i.e., the first block is assigned a constant weight `n', followed by `n-1' for the next block, and so on. The algorithm in Fig. \ref{6} is explained using the decaying step function in the following example.
\vspace{3pt}

\begin{example}\label{example1}
We consider a sample program with 10 MS gates (Fig. \ref{6}a) and divide it into two equal parts with five gates each. Any re-occurring gate in the first (second) block is assigned an edge weight of 2 (1). The function can be graphically shown in Fig. \ref{6}b. The first occurrence of the gate in the program irrespective of the block will be assigned a constant value equal to the number of gates in the program. The edge weights of the program are as follows: $wt(0,1) = 10+1+1+1=13$ (as the MS q[0],q[1] gate appears 4 times throughout the program, the three re-occurrences occur at the second block hence this block gets assigned edge weight of 1 ), $wt(1,2) = 10+1=11$ , and $wt(4,5) = wt(2,3) = wt(3,5) = wt(4,2) = 10$.
\end{example}

\subsubsection{Linear Decay Function}
One of the major drawbacks of using the step function becomes evident for programs with many gates where the blocks themselves contain a large number of gates with the same edge weight without considering the order in which the gates occur. For example, consider a program of 1000 2-qubit gates. For simplicity, we divide the program into 10 equal blocks with 100 gates each. Therefore, the same constant edge weight will be allotted for any re-occurrence for the 100 gates in a specific block. Within a span of these 100 gates, this function does not prioritize the earlier gate re-occurrences within the block. We circumvent this problem by using continuous decaying functions. Using a continuous linearly decaying function, we assign any re-occurrence of a gate a linearly reducing edge weight value as we go deeper into the program to prioritize the gates that re-occur at the start of the program. We varied \emph{a} (empirical parameter) between 0 to 1 to compare the number of shuttles with the Greedy mapper for various benchmark programs and random circuits (mentioned in Section IV). The edge weight in linear function can be modeled as:

\begin{equation*}
    W_{linear} = G - (a \times cnt)
\end{equation*}

Where $G$ is the number of gates, $a = 0.1$ (empirically determined) and $cnt$ varies between 0 and (number of gates in the program - 1).
\vspace{3pt}
\subsubsection{Exponential Decay Function}

The motivation behind using an exponential decay function is to get a sharp decrease in the edge weight as we go deep into the program. We tune empirial parameter \emph{a} to experimentally determine an optimal solution for the least number of shuttles across all benchmark programs. The edge weight can be modeled as 

\begin{equation*}
    W_{exponential} = G \times a^{-cnt/G}
\end{equation*}
Where $a = 2$ (empirically determined). Other definitions remain same as before.

\subsection{Penalized Linear Decay Function}
All the three functions mentioned above assign an edge weight (though reduced) as we go deeper into the program. 
However, in penalized approach, we assign a negative weight to a gate re-occurrence after a certain point in the program i.e., we penalize any re-occurrence of a gate after a certain point in the program. Intuitively, the qubits with gates occurring at a later stage in a program will need to shuttle out more often if mapped predominantly by the weights determined by the early occurring gates. Hence, such qubits would require a different mapping negating the weights determined by the early occurring gates. 
The proposed function also considers the program parameters such as, the total number of gates, depth of the program, number of qubits, and symmetric/asymmetric nature of the program. \emph{We define a symmetric program to have a fixed repetitive occurring pattern of gates throughout the program}.
The edge weight is modeled as
\begin{equation*}
    W_{P} = G - (S \times Q \times D/G) \times cnt
\end{equation*}

Where \emph{G} is the number of gates, \emph{Q} is the total number of qubits, \emph{D} is the depth of the circuit, and \emph{S} is the symmetry factor. For symmetric circuits \emph{S} takes the value 0, else 1. We define a symmetric program as the one where each gate will have the same number of re-occurrences and at the same relative time in the program when compared to other gates.

\begin{figure}[htbp]
\centerline{\includegraphics[width= 3.5in]{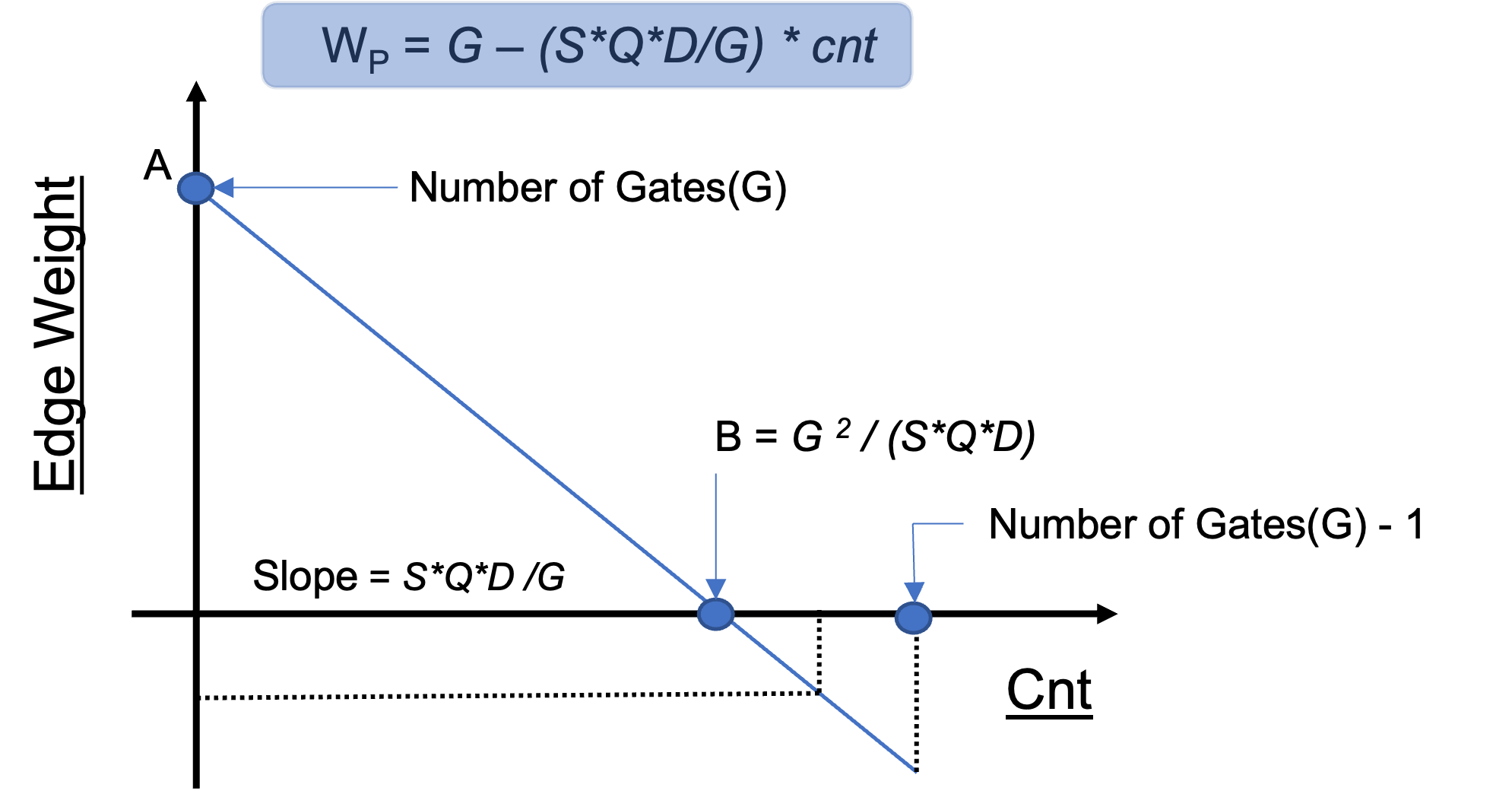}}
\caption{Graphical representation of a Linear Decay function for implementing the penalizing policy.}
\label{7}
\end{figure}

The edge weight allotment policy using this heuristic can be explained using Fig. \ref{7}. A program is divided into two parts by the point \emph{B}. Any re-occurrence of a gate before that point will be assigned a positive edge weight though small in magnitude compared to a preceding gate as per the function. Any re-occurrence of a gate post point \emph{B} in the program will be penalized by assigning a negative weight. The point \emph{B} for a program depends on an empirically determined \emph{function} of \emph{number of gates, depth, symmetry, and number of qubits.}

\section{Evaluation and Results}

\subsection{Experimental Setup}

\subsubsection{Hardware Model}
We use the linear TI hardware model as the one used in \cite{b23}-\cite{b24}. We consider the ``L6'' trap topology \cite{b23} where six traps are connected linearly. For each trap, initially, 15 ions can be loaded. 
\vspace{3pt}

\subsubsection{Benchmark Programs}
Considering the  baseline capability for 50-100 qubit NISQ
systems, we selected applications with 60-80 qubits and 500-4000 two-qubit gates \cite{b23}. Our benchmark suite includes circuits from Google’s supremacy experiment, quantum approximate optimization algorithm (QAOA), quantum Fourier transform (QFT), a quantum arithmetic circuit Square Root \cite{b23}. To exapnd the benchmark suite, we also test for 120 randomly generated circuits which covers a wide range of communication patterns, number of qubits, number of gates and depths.\\

\begin{table}[]
    \centering
    \caption{Reduction in the number of shuttles}
    \begin{tabular}{cccccccc}
    \hline
     Benchmark & Qubits & Gates & Greedy & This Work & $\Delta (\downarrow)$\\ 
     \hline
     SquareRoot & 78 & 1028 & 355 & 348 & 7\\
     Supremacy & 64 & 560 & 233 & 217 & 16\\
     QAOA & 64 & 1260 & 975 & 975 & 0\\
     QFT & 64 & 4032 & 196 & 196 & 0\\
     \hline
     \end{tabular}
    \label{tab:1}
\end{table}


\subsection{Results}

\subsubsection{Shuttle Reduction}

Table~\ref{tab:1} shows the reduction in the number of shuttle operations using the new heuristics compared to the Greedy mapper for the benchmarks circuits especially for square root and supremacy. For symmetric programs such as, QAOA and QFT circuits, our algorithms still show similar performance as that of Greedy as expected. The basic idea of prioritizing gates based on their re-occurrence becomes irrelevant for a symmetric program as each gate will have the same number of re-occurrences and at the same relative time in the program. Our program has a Symmetry factor to compensate for that. We also test our policy using 120 randomly generated circuits covering a wide variety of programs with Qubits varying from $\approx(60-75)$, number of gates between $\approx$ (900-2000), high depth, and diverse communication patterns. Table~\ref{tab:2} outlines the performance of different decaying functions and our policy (with Greedy policy being the baseline). We report the number of circuits for which the number of shuttles are reduced and increased, average reduction and increment in the number of shuttle operations and net reduction in the number of shuttles for the whole 120 random circuits. All three variants provide net reduction in average number of shuttles however, the penalizing function performs the best. Out of 120 circuits, we observe a decrease in the number of shuttles for 100 circuits with an average percentage decrease of $\approx$9\% for the penalizing weighting function. 
\vspace{3pt}
\subsubsection{Program fidelity improvement}
Shuttle operation increases vibrational energy $\bar{n}$,  of an ion-chain and degrades gate fidelity as per Eq.~\ref{eq:fidelity}. As our proposed policies reduces shuttles, it curbs motional mode resulting in the improved gate and overall program fidelity. For the random circuits, the exponential and linear decay functions provide 10\% and 6\% fidelity improvement, respectively (Table~\ref{tab:2}). The best gain of 1.41X on average (3.3X at best) is achieved with the penalized approach due to reduction in the net shuttle operations.
\vspace{3pt}
\subsubsection{Cost and Limitations}
All three variants provide net reduction in average number of shuttles. However, even for our best performing approach using the penalizing function we have seen a minuscule increase in shuttles for 20 random circuits. For symmetric programs such as, QAOA and QFT circuits, our algorithms show similar performance as that of Greedy mapping. Our mapping optimization for shuttle reduction does not show any increase in compilation time when compared to the greedy mapping policy as shown in Table~\ref{tab:3}.

\begin{table}[]
    \centering
    \caption{Analysis of 120 random circuits.\\ Baseline is the greedy mapper}
    \begin{tabular}{cccc}
    \hline
     Parameters & Exponential & Linear  & Penalized\\ 
     & $(a = 2)$ & $(a = 0.1)$ & \\
     \hline
     \# of ckts w/less shuttles & 73 & 77 & 100\\ 
     Avg. reduction in shuttles & 46.52 & 50.22 & 62.29\\
     \hline
     \# of ckts w/more shuttles & 47 & 43 & 20\\
     Avg. increase in shuttles & 38.86 & 43.60 & 29.55\\
     \hline
     Net reduction in shuttles  & 13.75  & 16.6  &  56.88 \\
     \hline
     Net \% reduction in shuttles & $\approx1.8\%$  & $\approx2\%$  & $\approx9\%$\\
     \hline
     Avg. net fidelity improvement & 1.1X  & 1.06X  & 1.41X\\
     \hline
     Max. fidelity improvement & 1.8X  & 2.67X  & 3.3X\\
     \hline
    
    \end{tabular}
    \label{tab:2}
\end{table}

\begin{table}[]
    \centering
    \caption{Compilation Time Overhead}
    \begin{tabular}{cccc}
    \hline
     Benchmark & Compile Time(sec) & Compile Time(sec)  & $\Delta (\uparrow)$ \\ 
     & $(Penalised)$ & $(Greedy)$ & \\
     \hline
     SquareRoot & 7.81 & 8.14 & -.33\\ 
     Supermacy & 3.26 & 3.55 & -.29\\
     QAOA & 19.18 & 19.25 & -.07\\
     QFT & 20.82 & 20.91 & -.09\\
   

     \hline
    
    \end{tabular}
    \label{tab:3}
\end{table}




\section{Conclusion}
In this paper, we present an efficient initial mapper for multi-trap TI quantum computers. Our technique achieves an average reduction of 9\% shuttles/program (with 21.3\% at best) for 120 random circuits and enhances the program fidelity up to 3.3X (1.41X on average) compared to the state-of-the-art Greedy mapper. The proposed policy also considers program-specific parameters such as, number of qubits, number of gates, and the depth for the initial qubit mapping, making it more holistic and efficient.

\end{document}